\documentstyle[12pt,aaspp4]{article}

\newcommand{\etal}{{\it et al.\,}}

\received{}
\revised{}
\accepted{}
\journalid{}{}
\articleid{}{}
\paperid{}
\ccc{}

\begin{document}

\title{Spectroscopic Observations of Optically Selected Clusters of Galaxies
	from the Palomar Distant Cluster Survey\footnote{Based on
	observations obtained with the Apache Point Observatory
	3.5-meter telescope, which is owned and operated by the
	Astrophysical Research Consortium.}}

\author{B. P. Holden\altaffilmark{2}}
\authoraddr{Department of Astronomy and Astrophysics, University of
Chicago, 5640 South Ellis Ave. Chicago, Illinois 60637}
\affil{Department of Astronomy and Astrophysics, University of
Chicago, 5640 South Ellis Ave. Chicago, Illinois 60637}
\affil{holden@oddjob.uchicago.edu}
\altaffiltext{2}{Visiting Astronomer, Kitt Peak Observatory, National
Optical Astronomy Observatories, which is operated by the Association
of Universities for Research in Astronomy (AURA), Inc., under cooperative
agreement with the National Science Foundation.}

\author{R. C. Nichol, A. K. Romer\altaffilmark{2}}
\authoraddr{Department of Physics, Carnegie Mellon University, 
5000 Forbes Ave. Pittsburgh, Pennsylvania 15213-3890} 
\affil{Department of Physics, Carnegie Mellon University,
5000 Forbes Ave. Pittsburgh, Pennsylvania 15213-3890}
\affil{nichol@andrew.cmu.edu, romer@andrew.cmu.edu}

\author{A. Metevier\altaffilmark{2}}
\authoraddr{Department of Astronomy and Astrophysics, 
University of California, Santa Cruz, California 95064} 
\affil{Department of Astronomy and Astrophysics, 
University of California, Santa Cruz, California 95064}
\affil{anne@ucolick.org}

\author{M. Postman\altaffilmark{2}} 
\authoraddr{Space Telescope Science Institute\altaffilmark{3},
3700 San Martin Dr., Baltimore, Maryland 21218}
\affil{Space Telescope Science Institute\altaffilmark{3}, 3700 San Martin Dr.,
Baltimore, Maryland 21218}  
\affil{postman@stsci.edu}
\altaffiltext{3}{The Space Telescope Science Institute is operated by
the AURA, Inc., under National Aeronautics and Space Administration
(NASA) Contract NAS 5-26555.}

\author{M. P. Ulmer} \authoraddr{Department of Physics and
Astronomy, Northwestern University, Dearborn Observatory, 2131 Sheridan
Road, Evanston, Illinois 60208-2900}
\affil{Department of Physics and Astronomy, Northwestern University,
2131 Sheridan Road, Evanston, Illinois 60208-2900} 
\affil{m-ulmer2@nwu.edu}

\author{L. M. Lubin} \authoraddr{California Institute of Technology,
105-24 Caltech, 1201 East California Blvd, Pasadena, California 91125}
\affil{California Institute of Technology, 105-24 Caltech, 1201 East
California Blvd, Pasadena, California 91125} \affil{lml@astro.caltech.edu}

\begin{abstract}

We have conducted a redshift survey of sixteen cluster candidates from
the Palomar Distant Cluster Survey (PDCS) to determine both the
density of PDCS clusters and the accuracy of the estimated redshifts
presented in the PDCS catalog (\cite{postman96} 1996).  We find that
the matched-filter redshift estimate presented in the PDCS has an
error $\sigma_z = 0.06$ in the redshift range $0.1 \le z \le 0.35$ based on
eight cluster candidates with three or more concordant galaxy
redshifts.

We measure the low redshift ($0.1 \le z \le 0.35$) space density of PDCS
clusters to be $31.3^{+30.5}_{-17.1} \times 10^{-6}\,h^3\,{\rm
Mpc^{-3}}$ (68\% confidence limits for a Poisson distribution) for
Richness Class 1 systems.  We find a tentative space density of
$10.4^{+23.4}_{-8.4}\times 10^{-6}\,h^3\,{\rm Mpc^{-3}}$ for Richness
Class 2 clusters.  These densities compare favorably with those found
for the whole of the PDCS and support the finding that the space
density of clusters in the PDCS is a factor of $\simeq5$ above that of
clusters in the Abell catalog (\cite{abell58} 1958; \cite{aco89}
1989).  These new space density measurements were derived as
independently as possible from the original PDCS analysis and
therefore, demonstrate the robustness of the original work.  Based on
our survey, we conclude that the PDCS matched-filter algorithm is
successful in detecting real clusters and in estimating their true
redshifts in the redshift range we surveyed.


\end{abstract}

\section{Introduction}

One of the focal points of modern observational cosmology is the
determination of the matter density of the universe.  Measurements of
the abundance of clusters of galaxies provide some of the strongest
constraints on this cosmological parameter.  For example, in a high
density ($\Omega_m=1$) universe, one would predict rapid evolution in
the number density of massive clusters with redshift, while in a low
density ($\Omega_m<<1$) universe, one would expect an almost constant
number density of such systems (\cite{lacey93} 1993; \cite{viana96}
1996; \cite{oukbir97} 1997; \cite{bahcall97} 1997; \cite{reichart99}
1999).

In the past, it has been difficult to measure the number density of
massive clusters from optically-selected samples of clusters because
previous samples have typically lacked a quantifiable selection
function ({\it e.g.} \cite{gunn86} 1986; \cite{aco89} 1989;
\cite{couch91} 1991). However, in recent years, the emergence of
optical cluster catalogs created with a completely automated selection
(\cite{lumsden92}; \cite{dalton92}; \cite{postman96} 1996 or P96;
\cite{lidman96} 1996) have now made it possible to use optically
selected catalogs for space density measurements.  For example, the
Palomar Distant Cluster Survey (PDCS; which is contained in P96) was
the first fully-automated, objectively-selected cluster catalog to be
based on deep CCD imaging data.  The cluster catalog was constructed
from a galaxy survey covering 5 deg$^2$ to V = 23.8 (3$\sigma$
completeness). The PDCS contains 79 cluster candidates selected via a
matched-filter algorithm that used both the magnitude and positional
data of the galaxies to find clusters.  This matched-filter is based
on a model for the galaxy distribution; a King profile for the
positional data and a Schechter function for the luminosity
distribution of the galaxies in the clusters.  The strength of any
observed correlation of the data with this matched-filter can be used
to measure how well the model matches the data, with the strongest
correlations being assigned to cluster candidates.  In this way, the
PDCS also generated an estimated redshift, a galaxy richness and other
parameters for each cluster candidate based on the best-fitting model.

Using the original catalog of 79 clusters, P96 discovered two
important results.  First, they determined that the space density of
PDCS clusters, for a given Richness class, was constant with redshift.
Second, they showed that the measured space density of PDCS clusters
was a factor of $\sim5 \pm 2$ above that seen in the local universe as
measured from the Abell catalog (see P96 for a discussion of the Abell
catalog space density).  These results have important implications and
have already been used to constrain cosmological parameters.  For
example, \cite{bahcall97} (1997) used the observed constant space
density of PDCS clusters with redshift, along with other samples of
clusters, to constrain both $\Omega_m$ and $\sigma_8$ (the variance of
density perturbations on cluster scales).  \cite{bahcall97} (1997)
found that the PDCS space density was consistent with a low $\Omega_m$
cosmological model.  These results illustrate the potential of
objectively-selected, statistical catalogs of clusters to constrain
cosmological models.  The work of \cite{bahcall97} (1997), however,
rests on several untested assumptions about the intrinsic properties
of the PDCS clusters {\i.e.} their M/L ratio, estimated redshifts {\it
e.t.c.}, some of which will be discussed in this paper.

Computer-based optical cluster finding algorithms such as the
matched-filter used to produce the PDCS will certainly play a major
role in the construction of the next generation of catalogs of
clusters of galaxies. This has already started to happen with surveys
such as DeepRange (\cite{postman98} 1998), the ESO Imaging Survey
(\cite{olsen99} 1999a; \cite{olsen99b} 1999b), the MDS deep cluster
sample (\cite{ostrander98} 1998) and the survey of \cite{zaritsky97}
(1997) to name but a few.  As the size and redshift range of the
catalogs increase, more detailed studies of the cluster evolution will
be possible providing tighter constraints on the values of $\Omega_m$
and $\sigma_8$.  Therefore, our group has embarked on a long-term
program to study the properties of the clusters discovered by the
matched-filter algorithm.

In this paper, we present the results of a redshift survey of a subset
of PDCS cluster candidates designed to test the estimated redshifts of
these cluster candidates as well as re-measuring the space density of
the PDCS catalog using spectroscopic redshifts.  We designed this
redshift survey to select a statistical subsample of PDCS cluster
catalog which is independent of most of the cluster parameters derived
by P96.  This approach minimizes the effect on our conclusions of any
systematic errors in those parameters.  In Sec. 2 of this paper, we
discuss how we selected the cluster candidates for this present
spectroscopic survey.  Sec. 3 describes the spectroscopic observations
and data reduction.  We then characterize the distribution in redshift
of the clusters in Sec. 4.  In Sec. 5, we discuss the implications of
our results and we conclude in Sec.6 with a summary of this paper.

\section{PDCS Cluster Candidate Selection}

Thirteen of the sixteen target clusters used in our spectroscopic
survey were selected from the subsample of PDCS clusters randomly
observed by the ROSAT satellite (as discussed in \cite{holden97}
1997).  These X-ray observations detected emission near six PDCS
cluster candidates and measured upper-limits for twenty-five PDCS
cluster candidates.  Three of the six PDCS cluster candidates found
associated with X-ray emission in \cite{holden97} (1997) are part of
the sample discussed in this paper.  The remaining three (PDCS 11, 12
\& 23) were taken from the rest of the PDCS.  

The main goals of our spectroscopic survey were to check the measured
space density of PDCS clusters and the accuracy of the estimated
redshifts.  Therefore, we selected our targets as independently as
possible of the derived parameters given by P96.  We selected targets
using only the net number of $V_4\ <\ 21$, $V_4-I_4\ >\ 1$ galaxies
(the subscript 4 refers to the 4-Shooter camera used to construct the
PDCS, see P96 for details on the filter system used and the resulting
galaxy catalog) within a 2\farcm 5 radius aperture of the
PDCS cluster candidates' position.  By taking this approach, we still
select clusters that potentially have a true redshift vastly different
from the PDCS estimated redshift.  Such clusters would have been
ignored if we had imposed a matched-filter estimated redshift cut-off.

We chose the above magnitude limit, color limit and angular size to
maximize the number of cluster candidates we could observe given the
limitations of our instrumentation.  The color criterion was chosen to
maximize the number of prospective cluster members over field
galaxies.  The median color of the field galaxies in the PDCS is
$V_4-I_4\simeq1$, while the cluster members are usually redder than
this limit (\cite{lubin96} 1996).  We note here that this color
selection does bias us against clusters dominated by blue galaxy
members.  The aperture size we chose was simply the field of
view of the Cryogenic Camera spectrograph, one of the two instruments
we used.

To compute the number of galaxies in our aperture, we filtered the
PDCS galaxy catalog removing all galaxies fainter than our magnitude
limit and bluer than our color limit.  We then counted the number of
galaxies within the 2\farcm 5 radius aperture centered on each PDCS
cluster candidate's position.  We subtracted the expected number of
field galaxies from the number of galaxies in the aperture to obtain
an estimate of the net number of cluster candidate galaxies per
aperture.  The expected number of galaxies was computed for each of
the four different PDCS fields (00$^{\rm h}$, 02$^{\rm h}$, 09$^{\rm
h}$ and 13$^{\rm h}$) separately.  The surface density of field
galaxies was measured by finding the total number of galaxies in our
filtered catalog not within the 2\farcm 5 radius of a PDCS cluster
candidate and then dividing that number by the area covered by the
catalog but excluding the area within 2\farcm 5 of PDCS cluster
candidates.

We ranked the 31 PDCS cluster candidates from the analysis of
\cite{holden97} (1997) based on the net number of cluster galaxies as
calculated above.  We then observed the PDCS cluster candidates in
descending order on this list.  In this way, we ensured that any
partial subset of this sample of 31 would be complete.  We eventually
observed the top thirteen of these 31 candidates and these thirteen
clusters are listed in Table \ref{pdcsobs}.  We observed three other
clusters (PDCS 11, 12 \& 23) that are not part of this complete sample
but were nonetheless selected in the same way (and are also listed in
Table \ref{pdcsobs}.)  In Figure \ref{rich_v_z}, we plot the clusters
we observed (with stars) and the clusters we did not observe (with
circles) as a function of their PDCS estimated redshift and estimated
richness.  Immediately one can see that most of the clusters we
observed possess a low estimated redshift and a high estimated
richness.  Also apparent are two clusters (PDCS 41 and 42) that have a
high estimated redshift and a low estimated richness.  These
clusters will be discussed in Sec 4.3.

\section{Spectroscopic Data and Reduction}

We spectroscopically observed a total of 130 galaxies in the direction
of the sixteen PDCS cluster candidates discussed above (Table
\ref{pdcsobs}). These observations were carried out on either the Kitt
Peak National Observatory (KPNO) Mayall 4 meter or the ARC 3.5 meter
telescope at Apache Point Observatory (APO).  Below we discuss the
process of creating slit masks for the KPNO observations and galaxy
selection for the observations performed on both telescopes.

\subsection{KPNO Masks and Observations}

We observed 122 galaxies towards fifteen of the sixteen PDCS cluster
candidates in our survey on the KPNO Mayall 4 meter using the
Cryogenic Camera with the 770 Grism (blazed with a peak transmission
at $\sim$ 6000\AA) or the 780-2 Grism (blazed with a peak transmission
at $\sim$ 7100\AA) using multi-object slit masks.  The observations
were carried out over six nights from April 29, 1997 to May 1, 1997
(KPNO1 on Table \ref{pdcsobs}) as well as December 19, 1998 and
December 21, 1998 (KPNO2 on Table \ref{pdcsobs}).  All of the
multi-slit masks were constructed using a constant slit width of
2\farcs 5 and a slit length of at least 14\arcsec.

For every cluster candidate, our aim was to observe a statistical
subsample of the galaxies that met our original color ($V_4-I_4\ >\
1$) and magnitude criteria ($V_4\ <\ 21$).  Therefore, to construct a
mask, we weighted each galaxy linearly based on its observed color
with the highest priority going to the reddest galaxies.  This was
done to maximize the number of galaxies with strong absorption
features typical of the old, metal-rich, ellipticals typically found
in the cores of clusters.  The magnitude of the galaxies played no
part in the determination of the weights as we did not want to bias
ourselves against fainter, and possibly higher redshift, galaxies that
met our magnitude limit.  To obtain the optimal use of the multi-slit
masks, we also included galaxies blue-ward of our color cut but these
galaxies were only used when a ``red'' ($V_4-I_4\ >\ 1$) galaxy could
not be targeted.  To achieve this, the ``blue'' galaxies were given
uniform weights one hundred times smaller than the smallest possible
weights given to the ``red'' galaxies.  One multi-slit mask was
constructed for each cluster candidate with the final galaxy selection
being performed automatically by maximizing the total weight for each
mask.  In addition to galaxies, each mask required an alignment star
($15<V<17$) which was critical for ensuring that our target galaxies
were aligned within the slit-lets.  In some cases, we were forced to
move the desired mask centers away from the original PDCS cluster
centroid to obtain alignment stars.  The worst case was PDCS 60 which
had to be shifted by 1\arcmin\, leaving only four slit-lets within the
2.5\arcmin\ aperture used for candidate selection.  The details of the
observing are summarized in Table \ref{pdcsobs}, including the
position of the center of the masks and the number of slit-lets 
per mask.

During our six nights at KPNO we encountered variable seeing and
transparency which resulted in only 77 useful spectra out of the 122
galaxies observed.  For each mask, we also obtained He, Ne, and Ar arc
lamp calibration spectra as well as quartz spectral flats at the same
zenith angle.  For a few of the longer integration times, we obtained
arc calibrations before and after the observation, though we did not
see any noticeable difference in the arc line positions.

All the multi-slit spectral data were reduced as outlined in
\cite{ellingson89} (1989) using the IRAF aperture spectra package,
APEXTRACT.  The spectral flat field observations were used to remove
the CCD response function.  We found tracing the spectrum using a
third order Legendre polynomial yielded the best results and we found
the background to be well fit by a constant.  We used
variance-weighted extraction to produce the one dimensional spectra.
These one-dimensional spectra were wavelength calibrated with the arc
calibrations.  The resulting spectral coverage was 4000\AA\ to 9000\AA\ in
pixels of 4.75\AA\ for the 770 Grism.  We measured a resulting
resolution of 14.9\AA\ using the width of the arc calibration lines
for the same Grism.  For the 780-2 Grism, the spectra covered 4500\AA\
to 9500\AA\ with at 4.5\AA\ per pixel with a resolution of
14.4\AA .  In Figure \ref{example_spectra}, we present some examples
of our spectra.

\subsection{ARC Data}

We observed eight galaxies for PDCS 02 and 05 on the ARC 3.5m
at APO with the Double Imaging Spectrograph (DIS) on October 4th and
October 26th, 1997.  All observations were carried out with a long
slit rotated to observe two galaxies at once.
The DIS instrument possesses a dichroic which splits the incoming beam
into a blue camera, covering a wavelength range of 3800\AA\ to
6000\AA\ with 6.3\AA\ per pixel, and a red camera, which covers a
wavelength range of 5000\AA\ to 9000\AA\ with 7\AA\ per pixel.  The
resolution of the red camera is measured to be 24.5\AA\ while the
resolution of the blue camera is 10.6\AA\ based on the full-width at
half-maximum of arc calibration lines.

At the time of the observations both DIS cameras had high read noise,
the blue CCD chip had a read noise of 15 e$^{-}$ while the red CCD
chip had 28 e$^{-}$.  The blue chip is comparable with the Cryogenic
Camera CCD chip in terms of read-noise and quantum efficiency so we
relied on data mostly from this camera for making the redshift
identifications of our target objects.  We used the data from the red
camera only to search for emission lines.

Our galaxy selection was done in a similar manner as discussed above
for the Cryogenic Camera.  As we were not using multi-slit masks, we
did not need to include any of the ``blue'' objects discussed above,
we simply sorted the ``red'' target galaxies ($V_4-I_4\ \ge\ 1.0$) in our
2\farcm 5 radius fields by magnitude and began by observing the
brightest objects.  PDCS 02 was also observed at KPNO with the CryoCam
spectrograph yielding a total of seven useful spectra (three from the
KPNO observing run and four from the ARC observing run).  Table
\ref{galaxydata} presents the data from both instruments.

We obtained spectral flats as well as He and Ar arc lamps calibrations
at the beginning of every night.  We tested the dispersion corrections
using sky lines and found them to be stable at the low resolutions we
used.  For some objects a small shift on the order of a half a pixel
was required.  Each spectrum was shifted appropriately before we
determined the redshift of the target object.

As with the Cryogenic Camera data, we used variance-weighted
extraction to create our one dimensional spectra.  We then combined
the red and blue spectra using SCOMBINE in IRAF to produce one
spectrum for each galaxy from 3800\AA\ to 9000\AA\ with 6.3\AA\ per
pixel, Figure \ref{example_spectra} shows one such spectrum.

\subsection{Galaxy Redshift Determination}

Overall, we obtained spectra for 130 galaxies but, as mentioned above,
not all these spectra yielded a redshift measurement. We present in
Table \ref{galaxydata} the 84 galaxies for which we obtained a
redshift, 77 from KPNO and 7 from the ARC 3.5m.  These redshifts were
determined either by using the IRAF command RVIDLINES to fit profiles
to emission lines or by using the RVSAO package (\cite{kurtz98} 1998)
to cross-correlate the spectra with absorption line galaxy templates.

Only six of the galaxy spectra have obvious emission lines.  This is a
likely result of the color selection criterion we imposed on the
targeted galaxies.  For PDCS 57 \#2, we have both an absorption and
emission line redshift which agree with each other within the
estimated errors.

All 130 galaxies were processed using the XCSAO procedure in the RVSAO
package.  The XCSAO command cross-correlates the spectrum with a
number of template spectra.  We used three template spectra provided
in the RVSAO package (M31, M32 and a sum of 1489 spectra called
fabtemp97, see \cite{kurtz98} 1998 for details on this last template).
Before cross-correlation, we removed the parts of the spectra around
bright sky lines, and the large Telluric absorption features, as many
of our spectra possessed large residual errors in these regions due to
inaccurate sky subtraction. This can be seen in Figure
\ref{example_spectra}.  We also trimmed the wavelength coverage of our
spectra before cross-correlating.  In general, for the KPNO spectra,
we trimmed off the first 500\AA\ and the last 500\AA .  For some
spectra, not all of the wavelength range could fit within the CCD
image, so we further trimmed the spectrum before cross-correlating.

In Table \ref{galaxydata} we provide the PDCS identification number,
the identification number assigned to that galaxy in the mask, the
J2000 coordinates of the galaxy, the galaxy's $V_4$ magnitude, the
$V_4-I_4$ color, the redshift with error, and the \cite{tonry79}
(1979) r value from the cross-correlation for absorption spectra.  We
consider all redshifts with r$>3$ to be secure measurements, as
recommended by \cite{kurtz98} (1998).  Emission line spectra are
marked with the letter ``e'' in the last column followed by the number of
lines used to fit the redshift.  We note here that the spectrum of
PDCS 61 \# 112 shows evidence for an active galactic nucleus, {\it
i.e.} one very wide emission line (see Figure \ref{example_spectra}).
We assign a tentative redshift to this active galactic nucleus
candidate assuming the emission lines is \ion{Mg}{2} at 2798\AA .
PDCS 57 \# 110 is listed twice in Table \ref{galaxydata}, one listing
for the emission line redshift and one listing for the absorption line
redshift.

\section{Cluster Redshift Distribution and Space Density}
 
As stated before, the main goal of our spectroscopic survey
was to determine the space density of PDCS clusters and to test the
estimated redshifts derived from the matched-filter algorithm.  Below
we discuss how we determined the global redshift of the observed PDCS
cluster candidates as well as discussing the probable error on the
PDCS estimated redshift.  We then discuss our derivation of the space
density of PDCS clusters via Monte Carlo simulations of our selection
process.

\subsection{Determining Cluster Redshifts}

Given the individual galaxy redshifts, we need to determine which PDCS
cluster candidates are likely to be real physical systems and assign
to them a global cluster redshift.  For our first estimate of the true
cluster redshift, we found the median redshift of all the galaxies
observed towards that cluster candidate.  In Table \ref{zs}, we list
for each of our sixteen PDCS cluster candidates this median redshift.
We also list our estimate of the ``best'' cluster redshift.  In cases
where we have three or more concordant cluster members within 1500 km
s$^{-1}$ of the median redshift, the median redshift of all of the
cluster galaxies is assigned as the ``best'' cluster redshift.  We
chose 1500 km s$^{-1}$ as it is approximately three times the typical
cluster velocity dispersion as given by \cite{girardi98} (1998).
However, for the seven cluster candidates for which we only have two
redshift measurements that agree to within 1500 km s$^{-1}$, we simply
average these two measurements and quote this as the ``best'' cluster
redshift.  PDCS 57 does not have a pair of galaxies within a 1500 km
s$^{-1}$ velocity separation, therefore we leave the ``best'' redshift
estimate blank.  The final column in Table \ref{zs} provides the
fraction of galaxy redshifts we measured to be within 1500 km s$^{-1}$
of the ``best'' redshift.  Also presented in Table \ref{zs} is an
error estimate on the ``best'' redshift. This error is only quoted for
the eight clusters with three or more galaxies in agreement.  The
error listed is the median absolute deviation of the galaxy redshifts
in km s$^{-1}$.  The value for PDCS 02 is quite high because of the
large number of redshifts (four out seven) that are not near the
median redshift, thus leading to a possibly large over-estimate of the
error for the cluster redshift.

P96 predicts that $>$ 10\% of all PDCS clusters are
likely false detections from Monte Carlo simulations. Therefore, it
is possible that one, or two, of our spectroscopically observed PDCS
cluster candidates is not a real, physical cluster.  This problem
becomes more acute when we note that seven of our sixteen observed
PDCS cluster candidates only have two galaxies close in redshift space
(within 1500 km s$^{-1}$ of each other).  To measure the frequency of
such close pairs of galaxies in the field population of galaxies, we
randomly sampled the Canada-France Redshift Survey (CFRS;
\cite{lilly95}; \cite{lefevre95}; \cite{hammer95}; \cite{crampton95}).
Using the same color and magnitude limits, adjusted for slight
differences in the filters used, we extracted random 
groups of galaxies from the CFRS database, regardless of their
angular distribution. Initially, we chose a size of three
galaxies as this corresponded to the lowest number of galaxy redshifts
we observed towards any one of our sixteen PDCS candidates (see Table
\ref{pdcsobs}).  From these simulations, we discovered that $\sim$
16\% of all groupings of three galaxy redshifts, extracted from the
CFRS, possess a pair of galaxy redshifts with a separation less than
or equal 1500 km s$^{-1}$ or less.  If we increase the grouping size
to four galaxy redshifts extracted at once from the CFRS, the
situation becomes worse, with a close pair (within 1500 km s$^{-1}$)
occurring $\sim$ 27\% of the time. For a grouping size of five
galaxies, we find a close pair of galaxies $\sim$ 41\% of the time.

If we increase our requirement to observing a triplet of
galaxies, all within 1500 km s$^{-1}$ of either side of the their
median redshift, then the above rates of such an occurrence, as
observed from the CFRS database, drop dramatically to $\sim$ 3\% for
a grouping size of five galaxy redshifts.  Using these numbers as a
guide, for the remainder of this paper, we will consider those eight
PDCS cluster candidates which possess three or more galaxy redshifts
within 1500 km s$^{-1}$ of the median redshift as detected clusters.

\subsection{Matched-Filter Redshifts}

We have used our measured cluster redshifts to test the matched-filter
redshift estimate given by P96.  In Figure \ref{zs_v_ph}, we plot the
estimated redshift versus our spectroscopically measured redshift as
well as the expected 45\arcdeg\ line for comparison.  In this plot, the
solid squares represent cluster candidates with three or more
concordant redshifts, while the open circles represent clusters with
only two redshifts in agreement.

To quantify the observed accuracy of the matched-filter redshifts, we
calculated the standard deviation of the difference between the
matched-filter redshift and the spectroscopically measured redshift.
If we include only the eight clusters with three or more concordant galaxy
redshifts, as discussed above, the standard deviation is 
$\sigma_z = 0.06$.

In Figure \ref{zs_v_ph} there are two significant outliers; PDCS 41
and 42.  For these two PDCS cluster candidates, the galaxies we have
spectra for are near the cluster centroid but are significantly
brighter in magnitude than we would expect for any cluster members at
the estimated redshift of these two cluster candidates (see Figure 9
of P96 which shows that for $z\simeq0.8$ clusters, one would not expect
to see a single cluster member brighter than $V_4=21.5$).  For these
two cluster candidates therefore, we have either found a low-redshift
group masquerading as a high redshift cluster (with a significantly
different luminosity function), or we have found a chance
superposition of a pair of galaxies at low redshift (see the above
discussion concerning the probability of this happening).  In either
case, we can not rule out the presence of a high redshift cluster near
the estimated redshift for PDCS 41 and 42.

For completeness, if we include all fifteen of our PDCS cluster
candidates with at least two galaxies within 1500 km s$^{-1}$, the
standard deviation between the measured and matched-filter redshift
rises to $\sigma_z = 0.20$, which is in good agreement with the error
quoted by P96.  The increase in the error is mainly caused by the two
outliers discussed above (PDCS 41 and 42).  If we exclude PDCS
41 and 42, the error drops to $\sigma_z = 0.07$.

\subsection{The Space Density of PDCS Clusters}

We do not possess a volume limited sample of clusters because of the
selection criteria used in the creation of our sample.  Therefore, we
performed Monte Carlo simulations of our selection procedure to
estimate the range in redshift covered by our survey.  To compute the
probability of a cluster being selected for our survey, we constructed
model PDCS clusters and simulated the selection process outlined in
Sec 2.  We performed these simulations 10000 times to compute
the probability that we would detect PDCS clusters at different
redshifts.  These probabilities were then multiplied by the volume
element at that redshift (assuming a ${\rm q_o}$ and ${\rm H_o}$)
to produce an effective volume for a cluster with a given set of
properties.  These simulations can then be used to determine the
redshift range over which our survey is mostly complete and
hence statistically correct our incompleteness.

In our simulations, the clusters were modeled using a Schechter
luminosity function and a King model for the density profile.  The
parameters controlling the shape of the cluster luminosity function
and spatial profile were allowed to vary within their observed
distributions.  For the King profile, we used the median core radius
($r_c = 0.050$ Mpc h$^{-1}$) and slope ($\alpha = -1.36$) as estimated
by \cite{lubinpostman96} (1996) (see Table 2) for the PDCS. We also
used the variance estimates quoted by these authors.  For the
Schechter luminosity function, we assumed M$_{\star}$ = $-21.1$ in the
$V_4$ band and a slope of $\alpha=-1.1$, which are the values used in
P96. For the variance estimates of the luminosity function we used the
values quoted in \cite{colles89} (1989).  A random normalization
(denoted $\Lambda$ in P96) drawn from the actual values found in the
PDCS was used for each luminosity function. These normalizations,
however, were restricted to the ranges of Richness Class 0 ($20 \le
\Lambda_V \le 40$), 1 ($40 \le \Lambda_V \le 60$), and 2 ($60 \le
\Lambda_V \le 80$) type clusters.  We would like to note here that the
richness we use, $\Lambda$, has a redshift dependence (see Figure 19
of P96).  However, in the range of redshifts we are interested in,
$0.1 \le z \le 0.5$, the redshift dependence is quite mild so the
above ranges of $\Lambda_V$ reproduce the corresponding Abell Richness
Classes (see Fig. 17 of P96).

For each cluster, we assumed that the luminosity function only
contains galaxies redder than our color criterion, $V_4 - I_4 > 1$.
This assumption is not strictly correct, however, the majority
($\gtrsim$ 80\%) of all galaxies in PDCS cluster candidates are redder
than our color criterion (see Figure 7 of \cite{lubin96} 1996).

We note here that the apparent magnitudes of our simulated galaxies
depend not only on the cosmology and the luminosity function but also
on the k-corrections.  We used the k-corrections of an elliptical-type
galaxy (the nuclear bulge of M31) from \cite{coleman80} (1980) with
linear interpolation between the tabulated values.  We chose an elliptical
galaxy spectral energy distribution as this is most consistent with
the red galaxies found in the cores of these rich clusters.  We
used two values of the deceleration parameter, ${\rm q_o}$, 0.5 and
0.1. All of our results are computed with a Hubble Constant of ${\rm
H_o}$ = 100 km s$^{-1}$ Mpc$^{-1}$.

For each simulated cluster, we computed the expected number of cluster
galaxies to fall within a 2\farcm 5 aperture and above a magnitude
limit of $V_4\ =\ 21$.  We computed a Poisson deviate from this
expectation value.  We then computed another Poisson deviate using the
average number of background galaxies meeting our galaxy selection
criteria that fall within the 2\farcm 5 aperture as the expectation
value.  The sum of these two deviates is the simulated number of
galaxies.  From this number, we subtract the surface density of field
galaxies.  This difference is the net number of galaxies in our
aperture.

We performed the above simulation 10000 times for a given
Richness Class at each redshift interval of 0.01 out to $z=0.6$.  At
each redshift, we computed the percentage of times the net galaxy
count exceeded 2.5 galaxies, (the smallest net galaxy number listed in
Table 1).  This percentage provided an estimate of the probability of
such a cluster being included in our observing sample.  These
probabilities are summarized in Figure \ref{rc} for Richness Class 0
and 1 clusters as a function of redshift and for ${\rm q_o}$ = 0.5
(solid lines) and ${\rm q_o}$ = 0.1 (dashed lines).

As can be seen from Figure \ref{rc}, all model clusters, at any
redshift, have a finite probability of being included in our sample.
This is the result of chance fluctuations in the background being
large enough to cause a cluster to be selected even though none of the
cluster members would actually fall within our magnitude limit.  This
is a plausible explanation for why PDCS 41 and 42 are included in our
sample even though they have estimated redshifts so high that none of
the cluster members should be brighter than our magnitude limit.  If
we assume that the field galaxy population follows a Poisson
distribution, we can calculate the expected number of times that a
fluctuation in the background would result in a high redshift cluster
falling in our sample, such as PDCS 41.  We calculated this would
happen 18\% of the time for such clusters.  This agrees quite well
with the results of our Monte Carlo simulations (see Figure \ref{rc})
and thus provides a check on our method.

Using the results of the above simulations, we can now compute the
probability of a cluster being selected for our survey.  However, the
above simulations also show that a there is a probability of a cluster
at any redshift being selected for our survey, regardless of whether
or not we can correctly determine the cluster's redshift.  Therefore,
we need to determine over what range we can successfully measure the
redshift of a cluster given our observational strategy.  To estimate
the probability of correctly determining a cluster's redshift and,
thus the range in redshifts we probed, we have combined the two
different simulations discussed previously to construct mock redshift
catalogs.

To simulate redshift distributions, we used the CFRS dataset
from Section 4.1 to construct a background galaxy population.  We used
the above cluster models to predict the number of observable cluster
members.  We then model the redshift distribution of a cluster as a
Gaussian with a standard deviation of 500 km s$^{-1}$.  For every
cluster candidate in our sample, we construct 10000 mock redshift
distributions consisting of both cluster members and random galaxies
from the CFRS catalog.  Each redshift distribution has the number of
redshifts set to match the total number of $V_4-I_4\ >\ 1$ galaxy
redshifts we successfully measured towards that cluster candidate.
For example, for PDCS 62 we successfully measured six redshifts,
however only five have a $V_4-I_4\ >\ 1$, thus our mock redshift catalogs
would only have 5 redshifts.  The magnitude limit of the mock redshift
distribution is equal to faintest galaxy observed for that cluster
candidate from Table \ref{galaxydata}.

By constructing mock redshift distributions and then applying our
criteria for determining a cluster redshift from Section 4.1, we can
determine the probability of detecting each cluster in our sample.  We
ran the simulations twice, once for the estimated redshift from the
original PDCS catalog and once using the ``best'' redshift from Table
\ref{zs}.  For PDCS 57, which does not have a ``best'' redshift
estimate, we only created these mock redshift catalogs for the
estimated redshift.  The results of these simulations showed that the
failure rate became high for clusters with $z>0.4$.  For example, if
PDCS 62 is a cluster at a redshift of 0.4665 (the redshift of the the
closest pair), then we find that we would only have a 21.5\% chance of
detecting a triplet of galaxies all within 1500 km s$^{-1}$ of the
actual cluster redshift, despite the high richness of the system.
These artificial redshift distributions show that we can successfully
measure the redshift of Richness Class 1 and 2 clusters in our sample
for $z \le 0.35$, while for Richness Class 0 systems
we must restrict our results to $z \le 0.3$.

The above simulations of artificial redshift distributions have shown
that we should be able to reliably calculate the surface and space
density of PDCS clusters below a redshift of 0.3 regardless of the
richness of these clusters.  For Richness Class 1 and greater we
can extend the range to 0.35.  To ensure that the clusters are truly
physical systems, we will restrict our analysis to those clusters with
three or more concordant redshifts (as discussed above) and that are
part of the complete sample of \cite{holden97} (1997).  These
restrictions leave the following clusters for our consideration: PDCS
02, 05, 34, 36, 38 and 40.  Two of these clusters are Richness Class
0, three are Richness Class 1, and one is a Richness Class 2 cluster.
If we restrict the redshift range to $0.1 \le z \le 0.3$, we have four
PDCS clusters.  Over the 1.85 deg$^2$ (see \cite{holden97} 1997 for
details on the area surveyed), these four clusters have a surface
density of 2.2$^{+1.7}_{-1.0}$ clusters deg$^{-2}$.  This surface
density is in agreement with the bottom panel of Figure 21 of P96 for
clusters with a $\Lambda > 20$, given our large error bars.

For the space density we need to compute the volume enclosed by the
redshift range over the 1.85 deg$^2$ surveyed by the sample .  To
compute this volume, we used the results plotted in Figure \ref{rc}
and multiplied the probability of selection for each cluster by the
volume element for each redshift interval (the simulations were
performed at redshift intervals of $\delta_z = 0.01$).  In other
words, instead of the usual Friedman-Robertson-Walker metric volume,
we compute the volume as \[ V(z) = \Omega \int_{z_1}^{z_2} \frac{p(z)
r^2 dr}{\sqrt{1-kr^2}} \] where $p(z)$ represents the probability of a
cluster being selected for our survey, $\Omega = 1.85 $deg$^2$, $z_1 =
0.1$, $z_2 = 0.35$ and $k=0$ for $q_o = 0.5$.  We find that the
computed space density of the three Richness Class 1 clusters (simply
$\frac{3}{V(z)}$) is $31.3^{+30.5}_{-17.1}\times 10^{-6}\,h^3\,{\rm
Mpc^{-3}}$ (${\rm q_o}$ = 0.5, 68\% confidence limits for a Poisson
distribution, see \cite{gehrels} 1986).  For the same value of ${\rm
q_o}$, P96 measured $20.2 \pm 8.1 \times 10^{-6}\,h^3\,{\rm
Mpc^{-3}}$ for Richness Classes 1 clusters.  In the redshift range
$0.1 \le z \le 0.35$, we also include PDCS 34, a Richness Class 2 cluster.
This single Richness Class 2 cluster implies a space density for
Richness Class 2 clusters of $10.4^{+23.4}_{-8.4}\times
10^{-6}\,h^3\,{\rm Mpc^{-3}}$.  

\section{Discussion}

Our survey had two goals, to test the accuracy of the matched-filter
estimated redshift, and to estimate the surface and space density of
PDCS clusters. We find that the matched-filter redshift estimate is at
least as accurate as discussed in the original PDCS paper.  Secondly,
we find that the cumulative surface density and the space density of
PDCS clusters at low redshifts agrees with the density found using the
original catalog.

Our worst case estimate of the standard deviation for the
matched-filter redshift measurements is 0.20 in z, the same error
estimate quoted in the original paper of P96.  However, if we restrict
our sample to those eight clusters with three or more redshift in
agreement, we find that the dispersion drops to 0.06 in redshift.
This error is quite small and is comparable to the errors on
photometric redshifts for individual galaxies (\cite{brunner97} 1997).
Our measurement of the error in the matched-filter redshift most
likely represents the best possible case for this algorithm.  The
majority of the clusters in our sample are at low redshifts compared
to the majority of clusters in the PDCS ($0.1 \le z \le 0.35$) and
therefore the cluster members are the easy to separate from field
galaxies.  Also, as the cluster galaxies are at low redshift they are
quite bright and thus, have the best measured magnitudes in the PDCS.
For the higher redshift clusters in the PDCS, the error in the
measured magnitudes of the cluster members should increase (as the
magnitudes will be fainter) and there will be a higher surface density
of field galaxies contaminating the cluster.  Both of these effects
will affect how well the matched-filter algorithm can estimate the
redshift of the cluster.  Nonetheless, the low apparent scatter in the
estimated redshift for low redshift PDCS clusters shows the power of
the matched-filter approach for characterizing cluster properties.
With the addition of photometric redshifts, this technique should
become even more accurate.

Our density measurements compare favorably with those found in the
original PDCS catalog and are a factor of approximately $7^{+7}_{-4}$
times higher than the Abell catalog cluster densities (see P96 and
references therein for Abell cluster densities), though with large
statistical errors.  One of the questions raised by the original PDCS
paper was why the PDCS has a space density of clusters of galaxies at
low ($z \le 0.3$) redshifts $\sim$5 times greater than that of the
Abell catalog.  Though we have not answered this question, we have
shown this is not because a large number of the PDCS cluster
candidates are false positive detections.  We independently estimated
the space density of PDCS clusters using only the clusters with three
or more measured redshifts within 1500 km s$^{-1}$ of the median, a
criterion that occurs $\sim$ 3\% of the time for triplets of field
galaxies as seen in the CFRS.  Therefore, our space density is based
on likely physical systems.

To further explore the type of clusters detected by the PDCS,
we have expanded our original survey to include more spectra per
cluster as well as more clusters.  This will allow us to measure
velocity dispersions and therefore, the masses of the PDCS clusters,
thus allowing us to directly compare with both the models of
\cite{press74} (1974), \cite{bahcall97} (1997) and \cite{reichart99}
(1999) and with the Abell catalog using such samples as the ESO Nearby
Abell Cluster Sample (\cite{katgert96} 1996) or the catalog of
\cite{girardi98} (1998).  By making comparisons with other samples, we
will be able to understand the discrepancy between the space density
of the PDCS clusters and those found in the Abell catalog.  By
comparing with the models, we will then continue the process of using
such a catalog to constrain models of cluster formation and evolution.
Nonetheless, the agreement between our measured space densities and
the densities quoted by the original PDCS is an important verification
of the PDCS cluster catalog.

\section{Summary}

The goal of our spectroscopic survey was to test the main results of
the original Palomar Distant Cluster Survey (PDCS) paper (P96).  We
achieved this goal by observing a subsample of PDCS clusters for
multi-object spectroscopy that was chosen as independently as possible
from the parameters of the original PDCS catalog.  We selected our
sample from those PDCS cluster candidates with the largest number of
$V_4<21$, $V_4-I_4\ > 1$ galaxies in the spectrograph field of view.
We chose these criteria to select likely cluster members that were
observable with the instrumentation available.  We obtained 130
spectra of galaxies in the direction of
sixteen PDCS cluster candidates.  Due to bad weather, only 84 of these
130 spectra yielded a secure redshift measurement.

We successfully obtained at least three redshift measurements for
every PDCS cluster candidate in our sample.  Using Monte Carlo
simulations of the Canada-France Redshift Survey, we have shown that a
secure cluster redshift can be obtained from such a small amount of
data, if one restricts the analysis to clusters with three of more
concordant redshifts within 1500 km s$^{-1}$ of the median redshift.
This only occurred $\sim$3\% of the time within our Monte Carlo
simulations of the Canada-France Redshift Survey.  Eight of the
sixteen PDCS clusters observed satisfy this requirement and lie in the
redshift range $0.1 \le z \le 0.35$.

If we restrict ourselves to this subsample of eight clusters, the
observed standard deviation between the spectroscopically measured
redshift for the clusters and the matched-filter estimated redshifts
(given by P96) is $\sigma_z=0.06$.  If we consider all fifteen PDCS clusters
in our sample with at least two galaxies at the same redshift, this
observed standard deviation rises to $\delta_z=0.20$, a value
consistent with the error quoted by P96.

Using detailed simulations of our target selection procedure, we have
measured the space densities of Richness Class 1 and 2 PDCS
clusters. We are in close agreement with the space densities quoted in
P96 {\it i.e.} that the observed space density of $0.1 \le z \le 0.3$ PDCS
clusters is a factor of $\sim5$ greater than that observed nearby from
the Abell Catalog.  However, we note that our measurements are
$\sim7^{+7}_{-4}$ greater than the Abell Catalog space density.  We
hope to build this on sample in order to investigate those properties
of PDCS clusters important for cosmological studies {\it e.g.}
redshift distribution, richness function, and mass function.  With
this information, will be able to answer the important question of
what sort of clusters the PDCS is finding.

\section{Acknowledgments}

We would like to thank Richard Kron for useful discussions on sample
selection, Francisco Castander and Marianne Takamiya for help with the
spectroscopic data reduction, and Constance Rockosi for helpful
comments on an early draft of this paper.  Finally, we would like to
thank Jim DeVeny of NOAO for his help in the mask preparation and instrument
setup.  This project was funded in part by NASA grant NAG5-3202.  BH was 
partially supported by the Center for Astrophysical Research in Antarctica, 
a National Science Foundation Science and Technology Center.

\epsscale{0.7}

\plotone{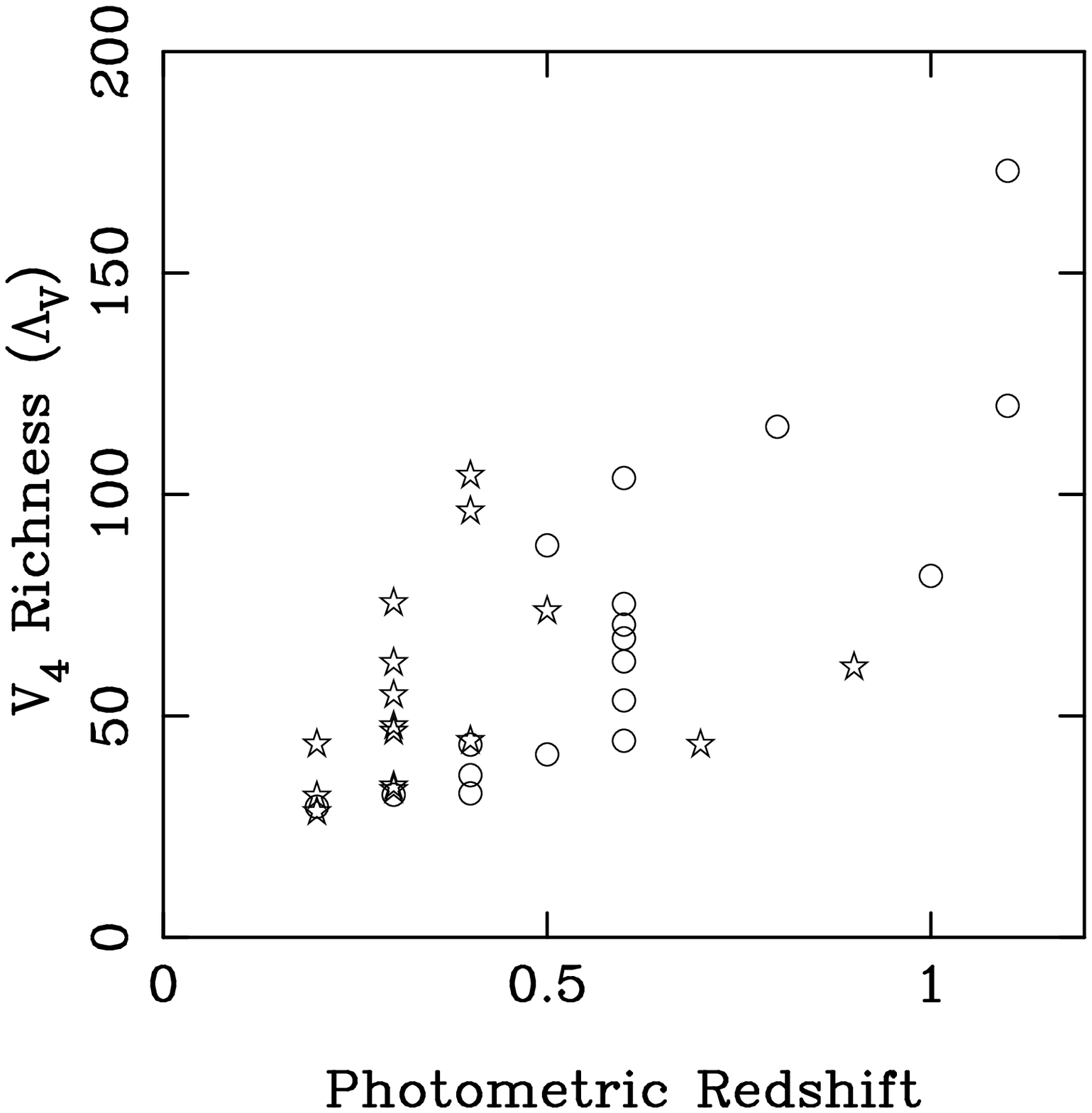}

\figcaption[holden.fig1.ps]{ The estimated matched-filter redshift versus the
observed V$_4$ richness for PDCS cluster candidates.  The stars
symbols represent the sixteen clusters observed and discussed herein while
the circle symbols are other PDCS in our sample that were not
observed.
\label{rich_v_z}}

\plotone{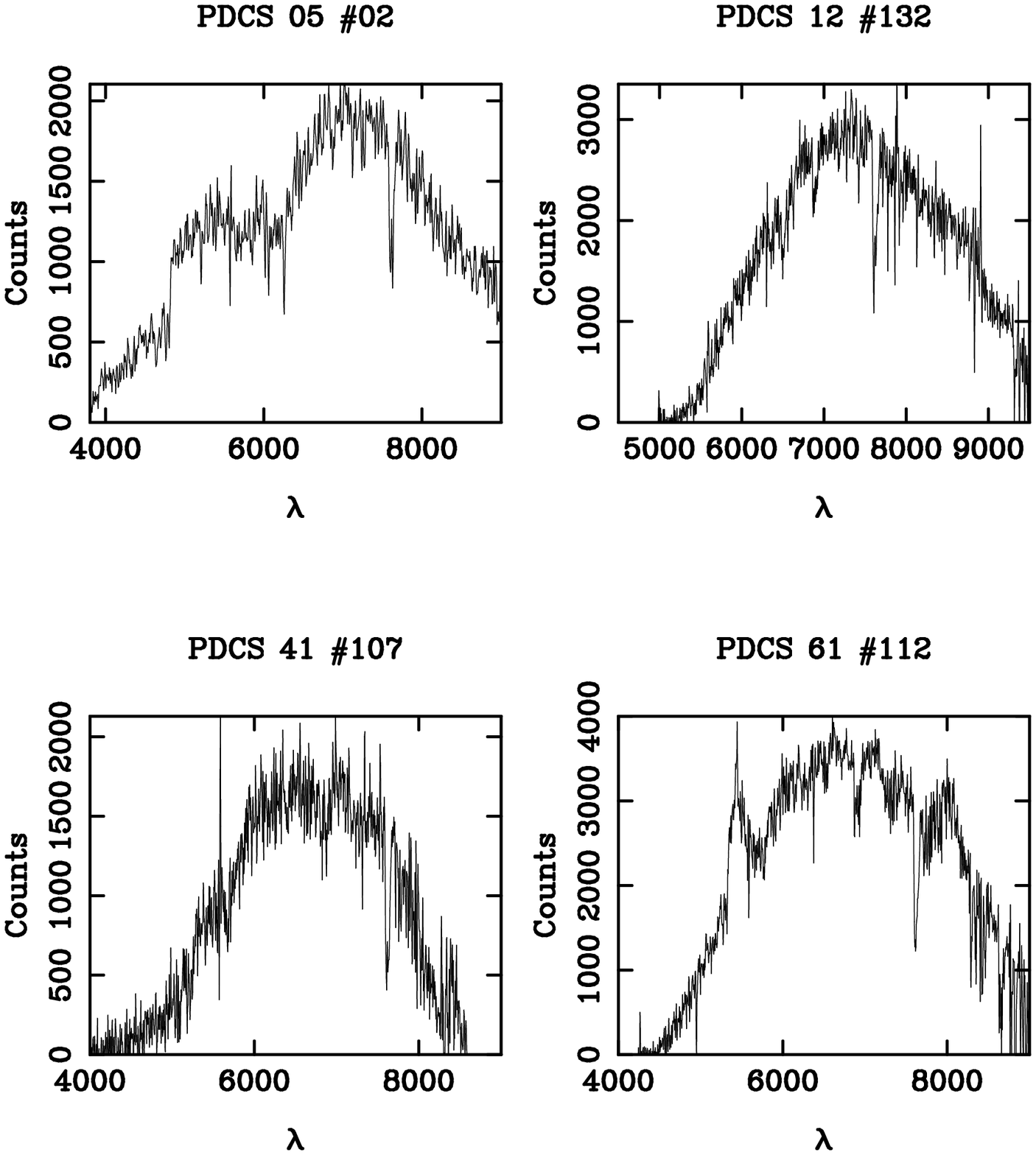}

\figcaption[holden.fig2.ps]{We plot representative spectra.  The upper
left-hand plot is PDCS 05 \# 2 (z=0.2112, r=6.05) which was observed
with the ARC 3.5m Double Imaging Spectrograph.  The spectrum shows
both the blue and red camera data.  The upper right-hand plot is PDCS
12 \# 132 (z=0.2585, r=4.67), a spectrum from CryoCam using the 780-2
Grism.  In the lower left-hand corner, we plot PDCS 41 \# 107
(z=0.3171, r-4.51), observed on CryoCam with the 770 Grism.  The
spectrum in the lower right-hand corner is a potential active galactic
nucleus found in our sample (PDCS 61 \# 112) which has a tentative redshift 
of 0.94 based on the one emission feature at 5437 \AA .
\label{example_spectra}}

\plotone{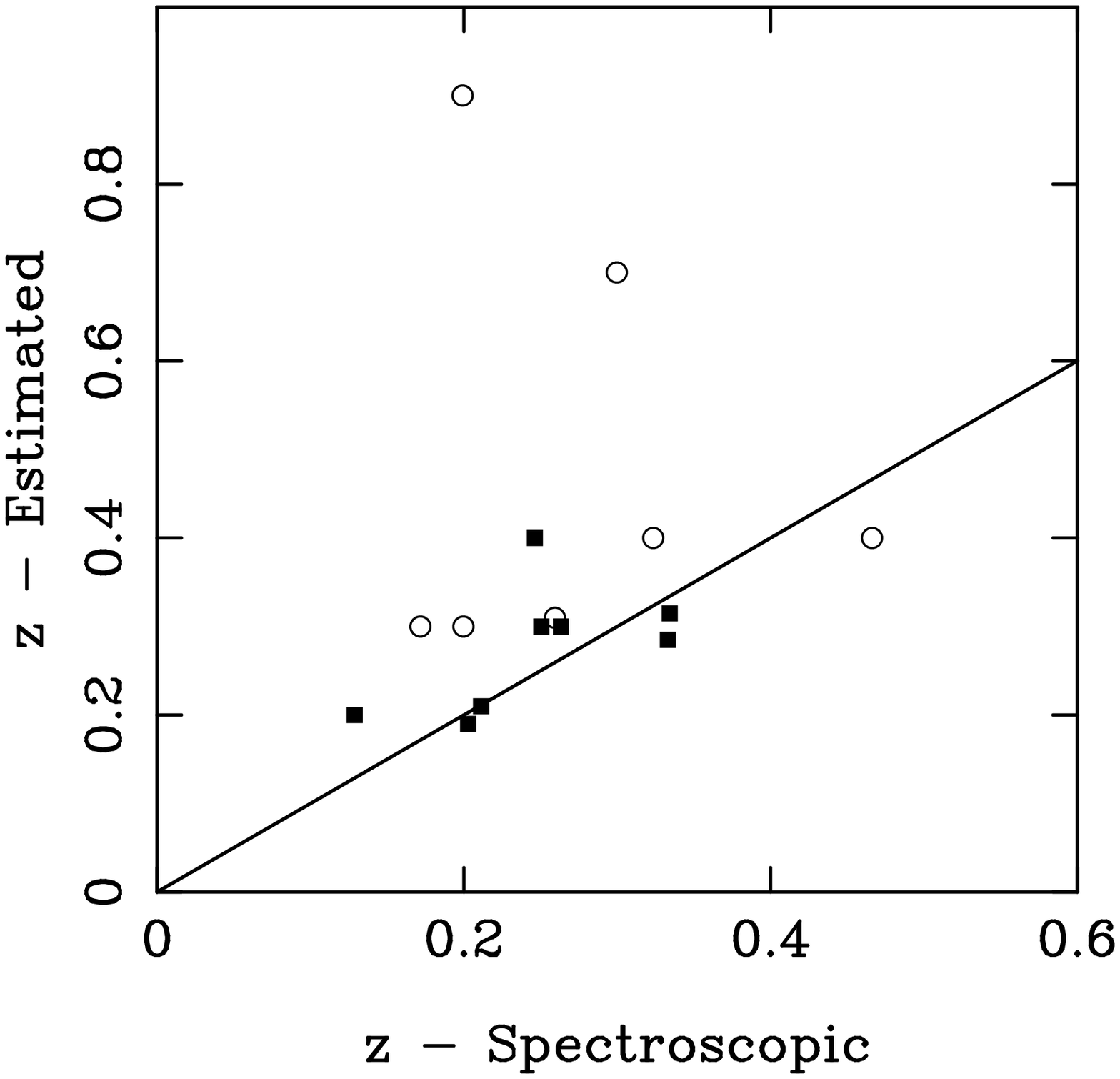}

\figcaption[holden.fig3.ps]{ We plot the matched-filter redshift
estimate of our PDCS cluster candidates versus their spectroscopically
observed redshift.  The filled squares represent clusters with 3 or
more coincident galaxy redshift measurements, while the open circles
represent clusters with only 2 coincident redshift measurements.  We
also plot a 45\arcdeg\ line for comparison.  Some of the data points
on the figure have been slightly offset from their actual estimated
redshifts to allow the reader to distinguish them.
\label{zs_v_ph}}

\plotone{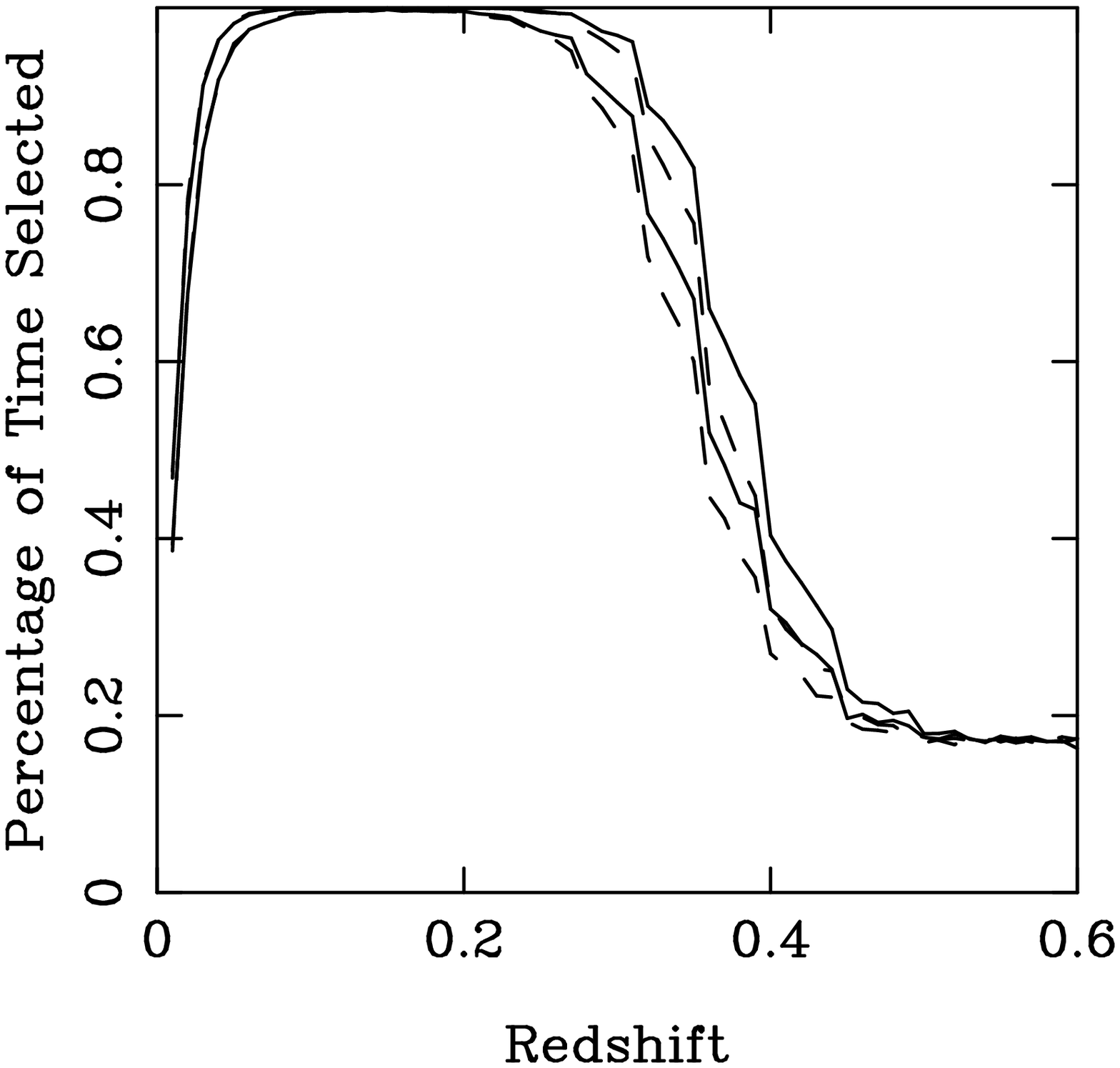}

\figcaption[holden.fig4.ps]{ Here we show the probability of
selecting a PDCS cluster for observation as a
function of redshift for ${\rm q_o} = 0.5$ (solid lines) and ${\rm
q_o} = 0.1$ (dashed lines).  The upper pair of lines represents the
probability of detecting a Richness Class 2 cluster, while the
bottom pair of lines is for Richness Class 1 cluster. 
\label{rc}}

\dummytable\label{pdcsobs}
\dummytable\label{galaxydata}
\dummytable\label{zs}

\end{document}